\documentclass{article}
\usepackage{amsmath}
\usepackage{latexsym}
\usepackage{amsfonts}
\usepackage{amssymb}
\usepackage{graphicx}
\begin{document}
\begin{center}
{\Large\bf Capillary Electrophoresis As a Fundamental Probe of Polymer Dynamics}
\\
George D. J. Phillies\\
Department of Physics\\
Worcester Polytechnic Institute\\
Worcester, MA 01609\\
phillies@wpi.edu\\
\end{center}

\begin{abstract}

Capillary electrophoresis has long been been recognized as a powerful analytic tool.  Here it is demonstrated that the same capillary electrophoretic experiments also reveal dynamic properties of the polymer solutions being used as the support medium.  The dependence of the electrophoretic mobility on the size of the probe and the properties of the matrix polymers shows a unity of behavior between electrophoresis and other methods of studying polymer properties.

\end{abstract}

\section{Introduction}

Following Svedberg's development\cite{svedberg1924a} of the ultracentrifuge, Tiselius and Svedberg\cite{tiselius1926a} demonstrated the experimental feasibility of electrophoretic separations of proteins and other polyelectrolytes.  Ultracentrifugation and electrophoresis are fundamentally similar. In each method, an external force is applied to individual macromolecules, causing them to migrate.  The migration rate is measured. For electrophoresis the migration velocity $v$ is related to the applied field $E$ and the electrophoretic mobility $\mu$ via
\begin{equation}
      v = \mu E.
      \label{eq:mobility}
\end{equation}
As seen below, $\mu$ in solutions of a neutral polymer is determined by variables including the probe size and the concentration and molecular weight of the matrix polymer.  With small applied fields, $\mu$ is a constant.  If the field strength is made large enough, $\mu$ becomes dependent on $E$.

The resolving power of an electrophoresis experiment is greatly increased if the simple bulk solvents studied by Tiselius are replaced with polymer solutions or gels. Electrophoresis in glass capillaries using a polymer solution support medium can rapidly separate double-stranded DNA molecules of molecular weight as large as $10^{3}$ base pairs with single-base-pair resolution\cite{albargh2000a}.  Gels and polymer solutions improve separations by obstructing convection.  They also act directly to improve the resolution of similar species.  Polymer solutions have differential effects on the electrophoretic mobility of homologous molecules, permitting separations that can not be accomplished with electrophoresis in simple fluids.

Modern research on electrophoresis is very substantially focused on improving the resolution and speed of separation procedures. Viovy\cite{viovy2000a} did compare electrophoresis of DNAs with particular models for polymer dynamics.  However, Viovy's thorough review was almost entirely concerned with electrophoresis through true cross-linked gels, rather than electrophoresis in polymer solutions as considered here.

The objective of this paper is to use the existing literature to show that electrophoresis can reveal interesting properties of polymer solutions.  The experiments on which this demonstration is based were designed to study separations, so conclusions on solution properties based on these experiments should be viewed as preliminary. This paper considers how $\mu$ depends on matrix polymer concentration $c$, probe and matrix molecular weights, and polymer topology; non-linear phenomena are briefly considered. Results here are representative of a larger body of analysis\cite{phillies2011a}.

\section{Dependence of $\mu$ on Probe and Matrix Properties}

This section considers electrophoretic studies, primarily on DNA fragments.  Molecular biology supplies DNA fragments having an extremely wide range of precisely-known molecular weights and geometric sizes.  Issues that can arise in optical probe diffusion studies with polymer-polymer binding and probe aggregation are not encountered with DNA fragments. Some naturally-occurring DNAs are huge, unentangled, unknotted rings; precision enzymatic single cuts transform these into linear DNAs of precisely the same molecular weight.  Synthetic methods for creating DNAs having exotic topologies, such as four-arm stars, have been demonstrated; their mobilities are discussed below.

Extensive measurements of electrophoretic mobility in polymer solutions were made by Barron, et al.\cite{barron1996a,barron1996b}.   Barron, et al., report electrophoretic mobilities of 16 DNA fragments, molecular weights 72-23130 bp (base pairs), in solutions of polydisperse polyacrylamide, hydroxyethylcellulose, and hydroxypropylcellulose at polymer concentrations up to 1.1\% w/w.  Prior to this work, it had been assumed based on particular models for polymer dynamics that only non-dilute polymer solutions could have a separatory effect on DNA fragments. Barron, et al.'s measurements were of great significance because they demonstrated that dilute polymer solutions are also useful for electrophoretic separations.

\begin{figure}
\includegraphics[scale=0.6]{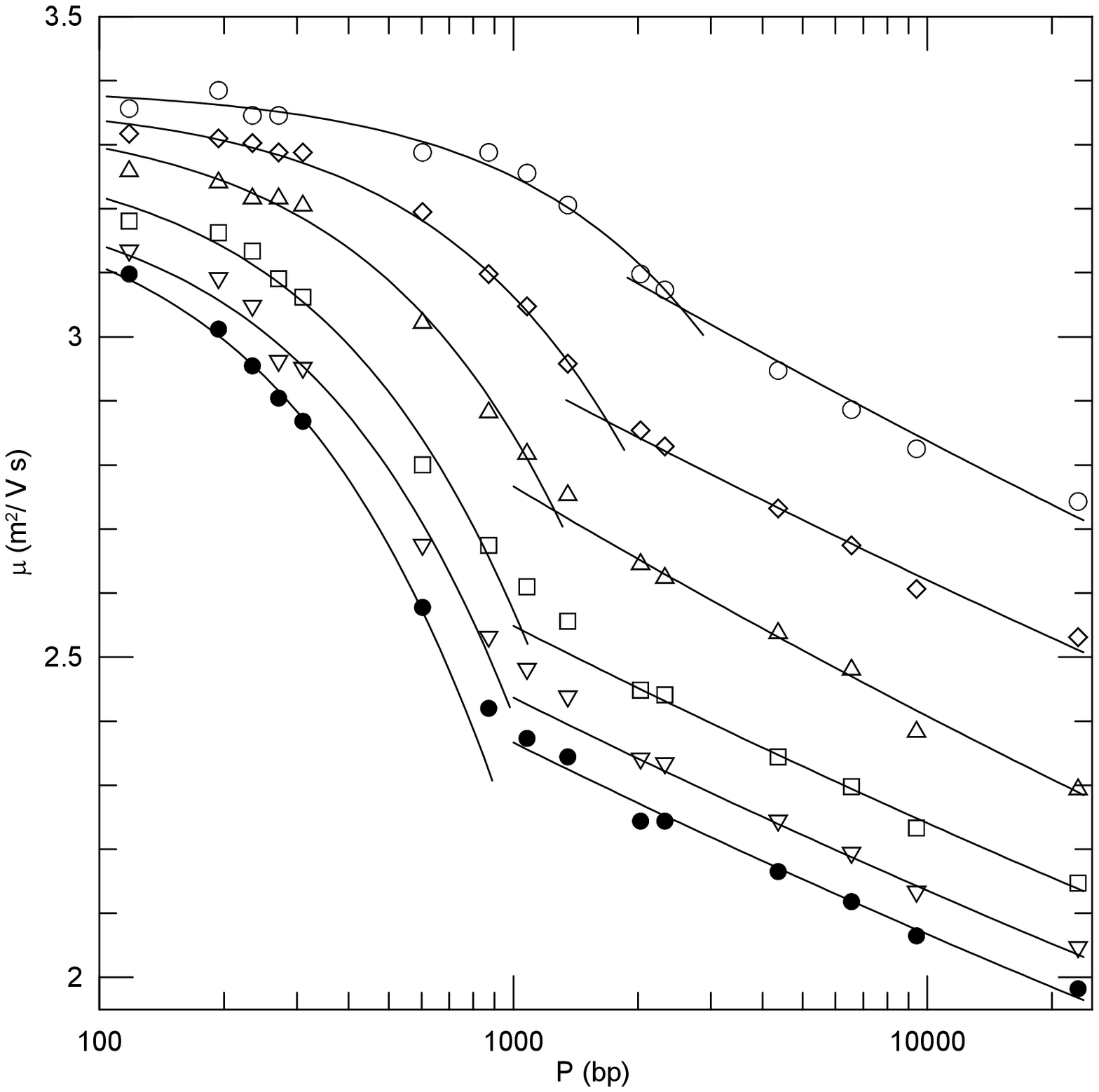} %
\caption{Electrophoretic mobility in 1 MDa hydroxypropylcellulose solutions at concentrations
($\circ$) 0.025, ($\lozenge$) 0.05, ($\vartriangle$) 0.1, ($\square$) 0.15, ($\triangledown$) 0.2, and ($\bullet$) 0.25 
fragments, as studied by Barron, et al.\cite{barron1996b} Lines are stretched exponentials and power laws in probe
molecular weight.}
\label{figurebarron1996b43RMm1}
\end{figure}

\ref{figurebarron1996b43RMm1} shows how $\mu$ depends on probe size, as measured by Barron, et al.\cite{barron1996b} at a series of concentrations.  The concentrations displayed here are a sample from a much larger number of concentrations reported in the original paper.   Barron, et al., estimate that their 1 MDa hydroxypropylcellulose matrix polymer had an entanglement threshold concentration $c^{*} \approx 0.09$ \% w/w, based on the concentration dependence of the solution viscosity, so the results shown in the Figure span nominally unentangled and mildly entangled ($c/c^{*} \approx 3$) solutions. As seen in the Figure, smaller probes have a stretched-exponential dependence
\begin{equation}
    \mu(P) = \mu_{1} \exp(-\alpha_{P} P^{\delta})
    \label{eq:mupSE}
\end{equation}
of $\mu$ on probe size. Larger probes show a power-law dependence
\begin{equation}
    \mu(P) = \bar{\mu} P^{\gamma}.
    \label{eq:mupPL}
\end{equation}
Here $\mu_{1}$ and $\bar{\mu}$ are nominal mobilities for a hypothetical tiny probe, while $\alpha_{P}$, $\delta$, and $\gamma$ are a scaling prefactor and two scaling exponents, respectively.  In this system, the transition between the stretched-exponential and power-law probe size dependences occurs for probes having 1000--2000 base pairs.

\begin{figure}
\includegraphics[scale=0.6]{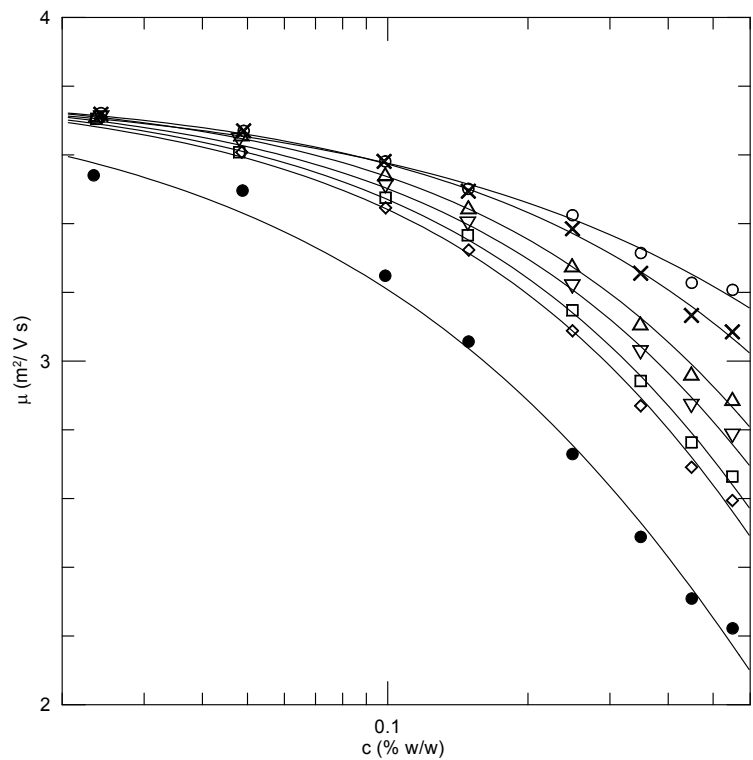}
\caption{Electrophoretic mobility of ($\circ$) 72, ($\times$) 118, ($\vartriangle$) 194, ($\triangledown$) 234, ($\square$) 271, ($\lozenge$) 310, and ($\bullet$) 603 base pair dsDNA restriction fragments in solutions of 1 MDa hydroxypropylcellulose, as measured by Barron, et al.\cite{barron1996b} Lines are stretched exponentials in c. }
\label{figurebarron1996aMm1}
\end{figure}

\begin{figure}
\includegraphics[scale=0.6]{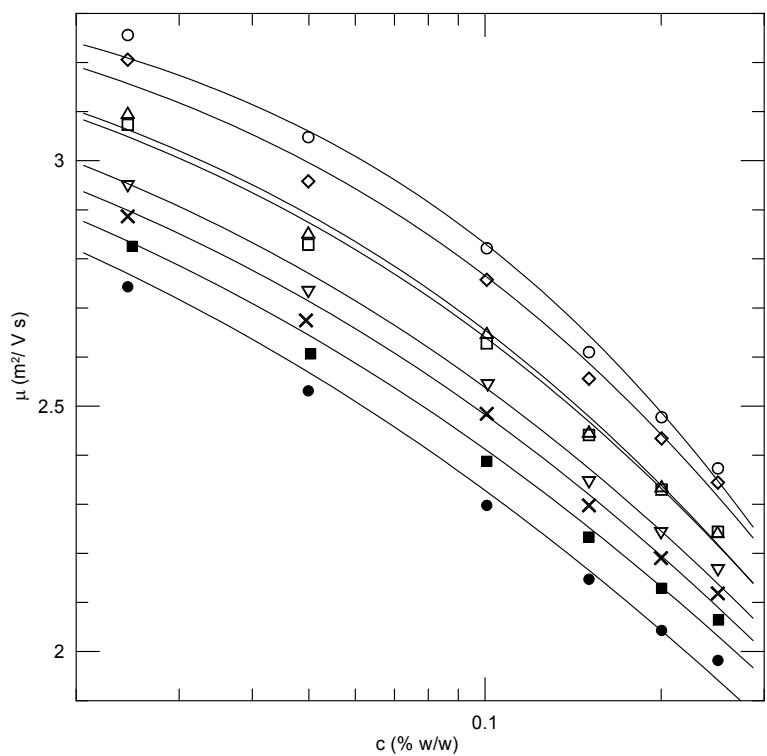}
\caption{Electrophoretic mobility of ($\bullet$) 23130, ($\blacksquare$) 9416, ($\times$) 6557, ($\triangledown$) 4361, ($\square$) 2322, ($\vartriangle$) 2027, ($\lozenge$) 1353 and ($\circ$) 1078 bp dsDNA restriction fragments in solutions of 1 MDa hydroxypropylcellulose, as measured by Barron, et al.\cite{barron1996b} Lines are stretched exponentials in c. }
\label{figurebarron1996b43Mm1}
\end{figure}

\ref{figurebarron1996aMm1} and \ref{figurebarron1996b43Mm1} show the concentration dependence of $\mu$. These are from the same study by Barron, et al., as seen in the prior Figure.  Now $c$ instead of $P$ is the abscissa, while point styles denote $P$ rather than $c$.   \ref{figurebarron1996aMm1} refers to small probes, for which $\mu(P)$ is a stretched exponential in $P$, while \ref{figurebarron1996b43Mm1} refers to larger probes, for which $\mu(P)$ follows a power law in $P$. $\mu(c)$ for each probe, small or large, follows to fair approximation a stretched exponential
\begin{equation}
    \mu(c) = \mu_{0} \exp(-\alpha_{c} c^{\nu}),
    \label{eq:mucfit}
\end{equation}
in $c$.  Here $\mu_{0}$ is the dilute-solution limit of $\mu$, while $\alpha_{c}$ and $\nu$ are a additional scaling parameters.

\begin{figure}
\includegraphics[scale=0.6]{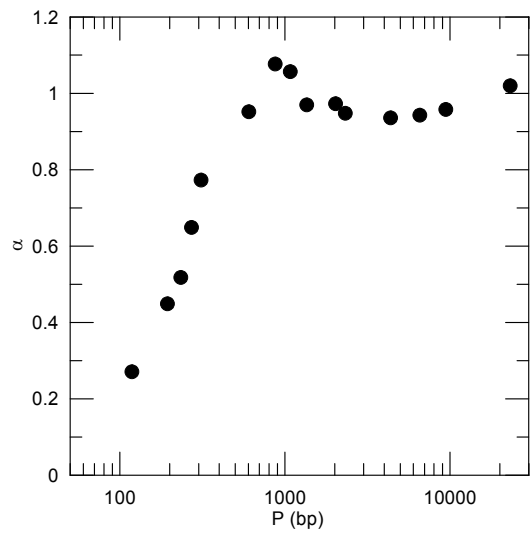}
\caption{Scaling prefactor $\alpha_{c}$ from \ref{eq:mucfit}, for DNA probes having various numbers $P$
of base pairs, as used to generate the smooth curves in \ref{figurebarron1996aMm1} and \ref{figurebarron1996b43Mm1}.}
\label{figurealphaPmuel}
\end{figure}

\begin{figure}
\includegraphics[scale=0.6]{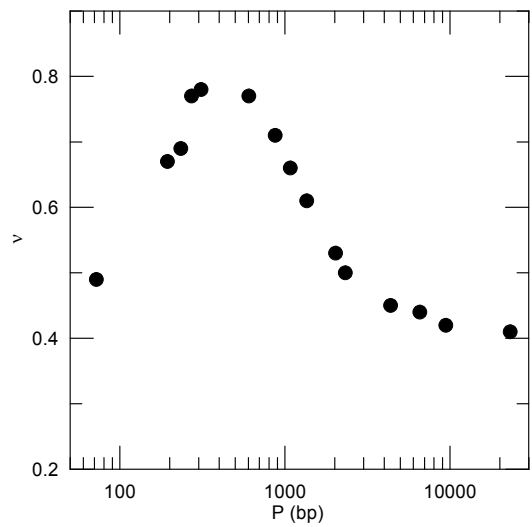}
\caption{Scaling exponent $\nu$ from \ref{eq:mucfit}, for DNA probes having various numbers $P$
of base pairs, as used to generate the smooth curves in  \ref{figurebarron1996aMm1} and \ref{figurebarron1996b43Mm1}.}
\label{figurenuPmuelMm1}
\end{figure}

\ref{figurealphaPmuel} and \ref{figurenuPmuelMm1} show the fitting parameters $\alpha$ and $\nu$ used to generate the smooth curves in \ref{figurebarron1996aMm1} and \ref{figurebarron1996b43Mm1}.  As seen in the Figures, $\alpha$ increases with increasing probe size, reaching a maximum near the 1000 bp threshold of the transition in the functional form of $\eta(P)$. For probes larger than 1000 bp, $\alpha$ depends at most modestly on $P$.  The scaling exponent $\nu$ also at first increases with increasing $P$, approaches $\nu \approx 0.8$ as the transition in $\mu(P)$ is approached, and appears to decrease toward a large-$P$ asymptote $\nu \approx 0.4$.  At large $P$,  $\alpha$ and $\nu$ thus have smooth dependences on $P$, dependences that are not the same for small and for large probes.

\begin{figure}

\includegraphics[scale=0.6]{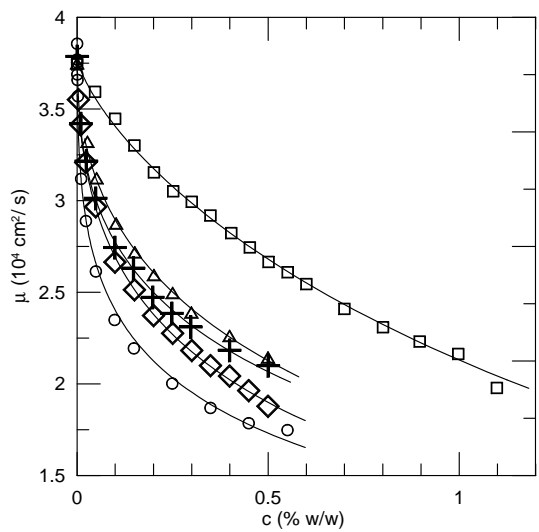}
\caption{ Electrophoretic mobility of 23130 bp dsDNA restriction fragments as functions of c for
($\circ$) 100, ($\vartriangle$) 300, and ($\lozenge$) 1000 kDa hydroxypropylcellulose solutions, as measured by Barron, et al.\cite{barron1996a}, showing the marked dependence of $\mu$ on matrix molecular weight. Lines are stretched
exponentials in polymer concentration.}
\label{figurebarron23e3bpMm1}
\end{figure}

\ref{figurebarron23e3bpMm1} shows the concentration dependence of $\mu$ for 23130 bp DNA in 100, 300, and 1000 kDa hydroxypropylcellulose solutions, again using measurements of Barron, et al.\cite{barron1996b}.  $\mu$ clearly has a substantial dependence on the molecular weight of the matrix polymer, $\mu$ decreasing as $M$ is increased.  These results are directly comparable to studies of Lodge and collaborators\cite{lodge1989a} on the diffusion of polymer tracer chains through polymer matrices having a range of $c$ and $M$, and are equally comparable to studies by Strelezky and Phillies\cite{streletzky1999a} on optical probe diffusion by polystyrene spheres in solutions of the 1 MDa hydroxypropylcellulose also studied by Barron, et al.\cite{barron1996b}. In all three sets of results, the measured transport coefficient depends strongly on matrix $M$.

\begin{figure}
\includegraphics[scale=0.6]{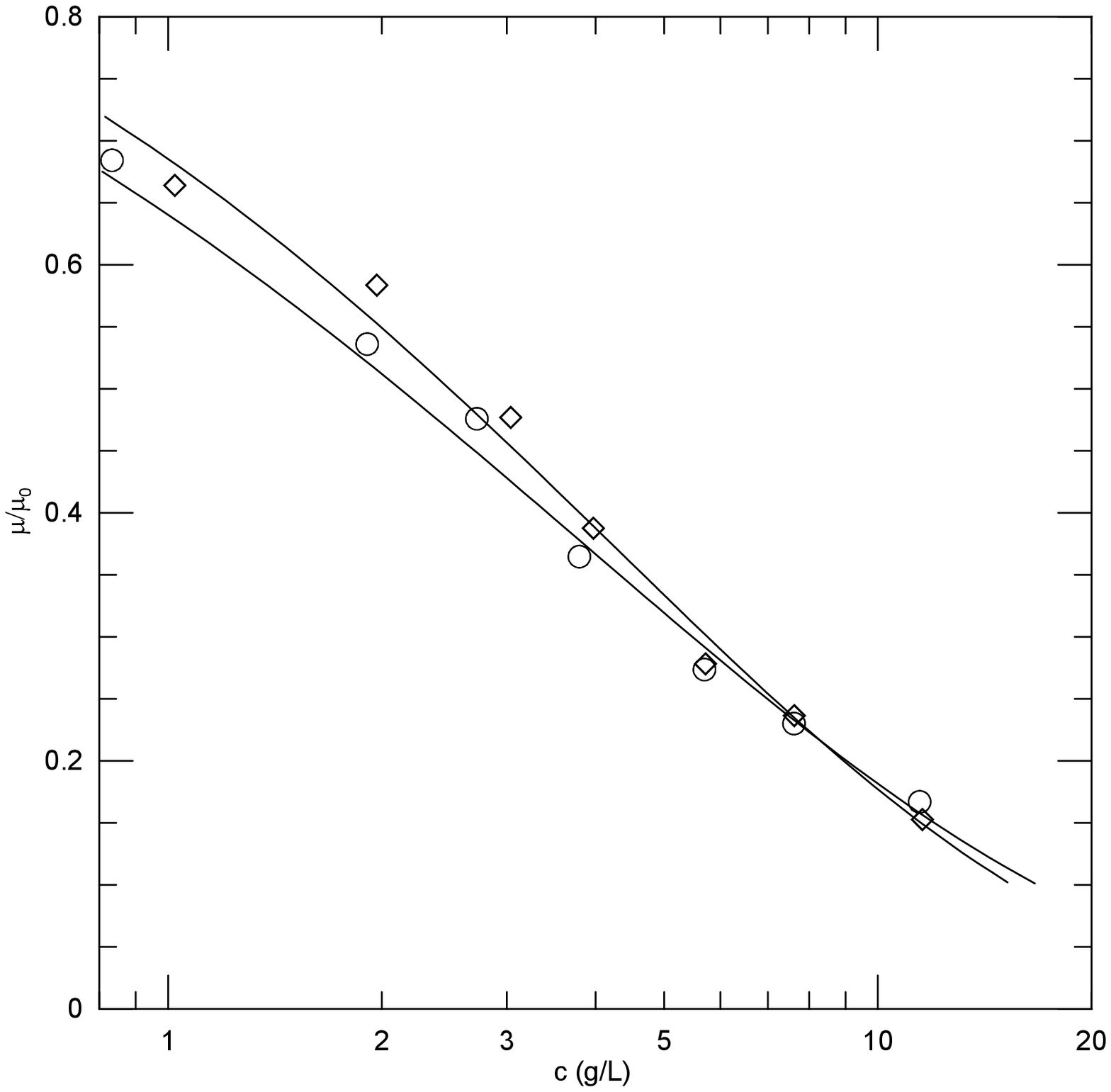}
\caption{Electrophoretic mobility of ($\lozenge$) 194 kbp synthetic four-arm star DNA and ($\circ$) 196 kbp
linear DNA in solutions of 1197 kda polyethylene oxide, as measured by Saha, et al.\cite{saha2004a}, showing
that $\mu$ is very nearly independent of probe topology in this system. Lines are stretched exponentials
in c.}
\label{figuresaha2004a3Mm1}
\end{figure}

The topology of the electrophoretic probes can be changed.  Modern biosynthetic methods allow the synthesis of DNAs having structures not found in nature\cite{heuer2003a}.  Despite their origins, such DNAs can be almost perfectly monodisperse, and can be prepared so that their arm or total molecular weight is very nearly the same as that of a matching linear DNA.  Saha, et al.\cite{saha2004a} used these methods to create 194,000 bp four-arm star DNA.  They compared the electrophoretic mobility of this star DNA with the mobility of 196,000 bp linear DNA in the same polymer solutions.  Representative measurements using a 1197 kDa polyethylene oxide matrix polymer appear as  \ref{figuresaha2004a3Mm1}.  $\mu$ of the star DNA and of the linear DNA are very nearly equal, showing that chain topology has almost no effect on chain mobility, at least for these matrix and probe molecular weights and the available range of concentrations.

One may also measure electrophoretic velocities of polystyrene spheres, denatured proteins, and microbes, as studied for example by Chrambach and Radko\cite{chrambach2000a}, Takagi and Karin\cite{takagi1996a}, Nakatani, et al.\cite{nakatani1996a}, and Armstrong, et al.\cite{armstrong1999a}.  Polystyrene spheres have historically been used extensively as probes of polymer dynamics in optical probe studies, in which spheres' diffusion coefficients were obtained.   The experiments here parallel the experiments sketched above on DNA electrophoresis, but the observed behaviors are not always the same.  Instead of quoting an electrophoretic mobility, one may for spheres quote the corresponding apparent (microscopic) viscosity $\eta_{\mu}$ opposing the electrophoretic motion. Here $\eta_{\mu} = \eta_{0} \mu/\mu_{0}$, the subscript $0$ referring to transport properties in the small-polymer-concentration limit.  Contrary to any expectation that the microscopic and macroscopic viscosities $\eta_{\mu}$ and $\eta$ should converge as probes are made larger and larger, the measurements of Radko and Chrambach\cite{radko1999a} show that $\eta_{\mu}$ is consistently less than $\eta$.  Furthermore, for polystyrene sphere probes $\eta_{\mu}/\eta$ in a given polymer solution decreases as the probes are made larger.  As a hypothesis, this behavior could be interpreted as stress thinning, in which the apparent viscosity is altered in proportion to the absolute force applied to the solution by the migrating probes, with larger spheres applying a greater force.

\section{Nonlinear Phenomena}

Modern research in polymer science has advanced from the study of linear viscoelastic properties to the study of non-linear viscoelastic phenomena.  Recently developed experimental techniques probe hitherto-inaccessible physical parameters.  Fourier transform rheology\cite{wilhelm1999a,neodhoefer2004a, fleury2004a,hyun2009a} considers a system subject to a shear at some frequency $\omega$, and measures responses at other frequencies $3\omega$, $5\omega$, \ldots. Medium and large angle oscillatory shear measurements determine the dynamic moduli also obtained by small-angle oscillatory shear, except using much larger maximum shear levels.\cite{hyun2009b}

Polymer solutions exhibit a variety of nonlinear response phenomena.  In addition to behaviors attendant to  non-zero normal stress differences\cite{hassager1984a,bird1977a}, behaviors are seen because polymeric fluids have relaxations on experimentally accessible time scales, as seen, e.g., in stress relaxation after single or multiple strains\cite{fukuda1975a,osaki1982a,archer2002a}.  As shown with modern technical means such as video microscopy, polymer solutions can also exhibit shear banding, in which the shear rate $d v_{x}/dy$ becomes dependent on $y$\cite{tapadia2006a,callaghan2008a}, and non-quiescent strain relaxation, in which an imposed strain is relaxed via large scale fluid motion rather than by molecular diffusion\cite{wang2006a,wang2008a}.

Here we catalog several nonlinear phenomena that have been observed in electrophoresing systems. As seen above, the probe-size-dependent difference between $\eta_{\mu}$ and $\eta$, as discussed above for electrophoresis of probe spheres, perhaps corresponds to stress thinning, stress thinning being a converse of shear thinning.  The bifunctional dependence of $\mu$ on $P$, distinct functions being required for small and for large probes, at least suggests non-linear behavior.  A true nonlinear regime, in which the electrophoretic transport coefficient changes with changing driving force, was demonstrated by Mitnik, et al.\cite{mitnik1995a}, who examined the dependence of $\mu$ on the applied field and other solution variables.  They found a low-field linear regime in which $\mu \sim E^{0}$, and an elevated-applied-field regime in which $\mu \sim E^{\gamma}$.  The nonlinear exponent $\gamma$ is larger at elevated polymer concentrations, but lies in the range $0.2-0.4$.  The field $E_{c}$, at which the transition occurs, declines with increasing probe size and apparently with increasing matrix concentration.  Radko and Chrambach\cite{radko1999a} report that electrophoresing polystyrene spheres can also be taken into a nonlinear transport regime, thereby changing the concentration dependence of $\mu$, by increasing the applied field.

Radko and Chrambach\cite{radko1999a} also describe a second nonlinear transport effect.  On the trailing side of migrating bands, they sometimes observed an extremely sharp subpeak.  No plausible species in solution could correspond to the subpeak, leading to the inference that the subpeak is created by nonlinear transport effects.  One notes that if the system could be manipulated so that the bulk of the migrating material were to be relocated into this nonlinear subpeak, the intrumental resolution and thus the number of species that might be resolved in the course of a single experiment might both be increased.

Finally, video microscopy has observed variations in the configuration of truly large DNAs during electrophoresis.  Ueda, et al.\cite{ueda1998a} found a novel DNA configuration, the "I" configuration, but only at elevated electrical fields. The crossover from low-field configurations to this larger-field configuration leads to a $\mu$ that changes with changing $E$.  Clearly this configurational change to the "I" configuration does not explain nonlinear effects that are also observed with electrophoresing polystyrene spheres.

From multiple lines of observation, electrophoresis thus reveals a probe-topology-independent, nonlinear transport regime at elevated $E$.  This observations are significant because the linear to nonlinear transition is highly reproducible.  Furthermore, the transition occurs in an experimentally convenient regime that is readily accessible to the experimental methods that reveal it.

\section{Remarks}

The objective of this short paper was to demonstrate that capillary zone electrophoresis, long employed as an analytic method, also has promise as a method for studying the dynamics of polymers in solution.  To find this method, a fundamental change of conceptual perspective was required. The critical conceptual inversion was to reverse the roles of the migrating species and the matrix polymer.  Instead of using the matrix polymer to enhance our understanding of the migrating charged species, the electrophoretic migration of charged species was used to enhance our understanding of the matrix polymer.

As demonstrations of this promise,  several classes of experiment taken from the published literature were examined, including (i) dependence of $\mu$ on matrix concentration and molecular weight, probe size, probe chain topology, and probe shape (rod or sphere), and (ii) electrophoretic phenomena that appear to correspond to nonlinear dynamic properties of polymer solutions.  A consistent phenomenology was uncovered.  The concentration dependence of $\mu$ in various systems follows a stretched exponential in $c$, precisely as found for other transport parameters.  The probe size dependence of $\mu$ shows a transition, showing a stretched exponential in $P$ for smaller probes and a power law for $P$ for larger probes.  The scaling parameters $\alpha$ and $\nu$ from $\mu(c)$ reflect the transition observed in $\mu(P)$, namely $\alpha$ and $\nu$ as functions of $P$ show one behavior for smaller $P$ and a different behavior for larger $P$.  The non-monotonic crossover behavior observed for $\mu(P)$ represents a novel phenomenon not previously seen with other experimental methods.  In particular, a corresponding transition ahs not been observed for probe diffusion in the same matrix polymer. The determination of $\mu(P)$ implicitly represent an important advantage of capillary electrophoresis over most other experimental techniques for the study of transport phenomena in polymer solutions, namely capillary electrophoresis can study probe motion by a large number of different probes at the same time, without any loss of instrumental resolution arising from the presence of multiple probes.

\pagebreak

\pagebreak

\pagebreak

\pagebreak

\pagebreak

\pagebreak

\pagebreak

\begin{thebibliography}{10.}

\bibitem{svedberg1924a} Svedberg, T.; Rinde, H.  {\em J. Am.\ Chem.\ Soc.}  {\bf 1924}, {\em 46}, 2677-2693.

\bibitem{tiselius1926a}   Tiselius, A.; Svedberg, T. {\em J. Am. Chem.\ Soc.} {\bf 1926}, {\em 48}, 2272-2278.

\bibitem{albargh2000a} Albarghouthi, M.~N.; Barron, A.~E. {\em Electrophoresis} {\bf 2000}, {\em 21}, 4096-4111.

\bibitem{viovy2000a} Viovy, J.-L.  {\em Revs.\ Modern Physics} {\bf 2000}, {\em 72}, 813-872.

\bibitem{phillies2011a} Phillies, G.~D.~J. {\em Phenomenology of Polymer Solution Dynamics}, Cambridge University Press {\em in press}.

\bibitem{barron1996a}  Barron, A.~E.;  Sunada, W.~M.; Blanch, H.~W. {\em Biotech.\ and Bioeng.} {\bf 1996}, {\em 52}, 259-270.

\bibitem{barron1996b}  Barron, A.~E.;  Sunada, W.~M.; Blanch, H.~W. {\em Electrophoresis} {\bf 1996}, {\em 17}, 744-757.

\bibitem{lodge1989a}  Lodge, T.~P.;  Markland, P.; Wheeler, L.~M.  {\em Macromolecules} {\bf 1989}, {\em 22}, 3409-3418.

\bibitem{streletzky1999a}  Streletzky, K.~A.; Phillies, G.~D.~J.  {\em Macromolecules} {\bf 1999}, {\em 32}, 145-152.

\bibitem{heuer2003a}  Heuer, D.~M.; Saha, S.; Archer, L.~A.
 {\em Electrophoresis} {\bf 2003}, {\em 24}, 3314-3322.

\bibitem{saha2004a} Saha,  S.;  Heuer, D.~M.; Archer, L.~A. {\em
Electrophoresis} {\bf 2004}, {\em 25}, 396-404.

\bibitem{chrambach2000a}  Chrambach, A.;  Radko, S.~P. {\em Electrophoresis}  {\bf 2000}, {\em 21}, 259-265.

\bibitem{takagi1996a}  Takagi, T.;  Karim, M.~R. {\em Electrophoresis}  {\bf 1996}, {\em 16}, 1463-1467.

\bibitem{nakatani1996a}   Nakatani, M.;  Shibukawa, A.;  Nakagawa, T.
{\em Electrophoresis} {\bf 1996}, {\em 17},  1584-1586.

\bibitem{armstrong1999a}  Armstrong, D.~W.;  Schulte, G.;  Schneiderheinze, J.~M.;  Westenberg, D.~J.
{\em  Anal.\ Chem.} {\bf 1999}, {\em 71},  5465-5469.

\bibitem{radko1999a}  Radko, S.P.; Chrambach, A.  {\em Macromolecules} {\bf 1999}, {\em 32},  2617-2628.

\bibitem{wilhelm1999a}  Wilhelm, M.;  Rienheimer, P.;  Ortseifer, M.  {\em Rheol. Acta} {\bf 1999}, {\em 38},  349-356.

\bibitem{neodhoefer2004a}  Neodhoefer, T.;  Sioula, S.;  Hadjichristidis, N.;  Wilhelm, M.
{\em Macromol. Rapid Communications} {\bf 2004}, {\em 25},  1921–1926.

\bibitem{fleury2004a}   Fleury, G.; Schlatter, G.;  Muller, R. {\em Rheologica Acta} {\bf 2004}, {\em 44},  174–187.

\bibitem{hyun2009a}  Hyun, K.;  Wilhelm, M.  {\em Macromolecules} {\bf 2009}, {\em 42},  411-422.

\bibitem{hyun2009b}  Hyun, K.;  Hoefl, S.; Kahle, S.;  Wilhelm, M. {\em J. Non-Newtonian Fluid Mech.} {\bf 2009}, {\em 160}, 93-103.

\bibitem{hassager1984a} Hassager, O. {\em J. Rheol.} {\bf 1984}, {\em 29},  361-364.

\bibitem{bird1977a}  Bird, R.~B.;  Armstrong, R.~C.;  Hassager, O. Dynamics of Polymeric Liquids. Vol. 1. Fluid Mechanics. {\bf New York: Wiley, 1977}.

\bibitem{fukuda1975a}  Fukuda, M.;  Osaki, K.;  Kurata, M.  {\em J. Polymer Sci.} {\em 13}, {\bf 1975},
1563–1576.

\bibitem{osaki1982a}  Osaki, K.;  Nishizawa, K.;  Kurata, M. {\em Macromolecules} {\bf 1982}, {\em 15},  1068-1071.

\bibitem{archer2002a}  Archer, L.~A.; Sanchez-Reyes, J.; Juliani.  {\em Macromolecules} {\bf 2002},  {\em 35}, 10216-10224.

\bibitem{tapadia2006a}  Tapadia, P.;  Ravindranath, S.;  Wang, S.-Q.  {\em Phys. Rev. Lett.} {\bf 2006},  {\em 96}, 196001 1–4.

\bibitem{callaghan2008a} Callaghan, P.~T.   {\em Rheol. Acta} {\bf 2008}, {\em 47}, 243-255.

\bibitem{wang2006a}  Wang, S.-Q.; Ravindranath, S.; Boukany, P.;  Olechnowicz, M.;  Quirk, R.~P.;
Halasa, A.;  Mays, J.  {\em Phys. Rev. Lett.} {\bf 2006}, {\em 97},  187801 1–4.

\bibitem{wang2008a}  Wang, S.-Q.  {\em J. Polymer. Sci: Part B: Polymer Physics}  {\bf 2008}, {\em 46},
2660-2665.

\bibitem{mitnik1995a} Mitnik, L.;  Salome, L.;  Viovy, J.~L.; Heller, C. {\em J. Chromatogr.\ A} {\bf 1995}, {\em 710}, 309-321.

\bibitem{ueda1998a} Ueda, M.; Oana, H.; Baba, Y.; Doi, M.; Yoshikawa, K. {\em Biophys.\ Chem.} {\bf 1998}, {\em 71}, 113–123.

\end{thebibliography}
\end{document}